\documentclass[notitlepage,nofootinbib,superscriptaddress,showpacs,preprintnumbers,amsmath,amssymb,prd,12 point]{revtex4-1}
\usepackage{epsfig,subfigure}
\usepackage{graphicx}
\usepackage{bbm}
\usepackage{enumerate}
\usepackage[colorlinks=true,citecolor=blue]{hyperref}
\draft 
\begin{document}
\author{S. Dev}%
\email{sdev@associates.iucaa.in}
\affiliation{Department of Physics, School of Sciences, HNBG Central University, Srinagar, Uttarakhand 246174, INDIA.}%
 \author{Desh Raj}%
\email{raj.physics88@gmail.com}
 \affiliation{Department of Physics, Himachal Pradesh University, Shimla 171005, INDIA.}%
\title{New Hybrid Textures for Neutrino Mass Matrices}
\begin{abstract}
We perform a systematic investigation of the texture structures of Majorana neutrino mass matrix $M_{\nu}$ having two texture zeros and an equality between two nonzero matrix elements, in the light of recent neutrino oscillation data. Among forty-two possible textures, it is found that only eight textures are compatible with the current experimental data at 3$\sigma$ C.L. Out of these phenomenologically viable textures, six follow normal mass ordering while remaining two satisfy the inverted mass ordering of neutrino mass spectrum. In the numerical analysis, we carry out a scan over the possible space of all viable patterns. We present the implications of each allowed patterns for three mixing angles (solar, reactor and atmospheric), leptonic CP-violation, neutrino mass scale and the neutrinoless double beta decay indicating strong correlations between oscillation parameters. The symmetry realization of one of the viable textures is also presented.   
\end{abstract}
\maketitle
\section{Introduction}
Since the discovery of neutrino oscillation by Super-Kamiokande experiment \cite{super} in the year 1998, considerable efforts have been put towards determining the structure of the neutrino mass matrix ($M_\nu$). In this direction, the neutrino oscillation parameters such as three neutrino mixing angles (solar, atmospheric and reactor) and two mass-squared differences ($\Delta m^{2}_{21}$ and $|\Delta m^{2}_{31}|$) have been measured with fair precision by various neutrino oscillation experiments. The experimental data from recent neutrino oscillation experiments hint towards a non-maximal atmospheric mixing angle \cite{t23} and the Dirac-type CP-violating phase near $270^\circ$ \cite{nova}. 
 Apart from these significant achievements, many other parameters like leptonic CP-violation, nature of neutrinos (Dirac or Majorana), absolute neutrino mass scale, the neutrino mass ordering and the origin of lepton flavor structure still remain unresolved issues. The neutrino mass matrix which encodes the neutrino properties contains many unknown physical parameters. The phenomenological models based on Abelian or non-Abelian flavor symmetries can result in specific textures of neutrino mass matrix with reduced number of independent parameters. Based on these approaches, several popular models such as texture zeros \cite{texture1,texture2}, vanishing cofactors \cite{minor}, equalities among elements/cofactors \cite{equal}, hybrid textures \cite{hybrid,liu} etc. are quite successful in explaining the presently available neutrino oscillation data. Hybrid textures  particularly one texture/cofactor equality with one texture/cofactor zero in $M_{\nu}$ have been studied extensively in the literature \cite{hybrid}.\\ 
In the present work, we investigate a specific class of neutrino mass matrices which contain two texture zeros and a texture equality between two nonzero elements in the flavor basis where the charged lepton mass matrix is diagonal. The main motivation of this work is as follows:
\begin{itemize}
\item From a phenomenological viewpoint, models of the neutrino mass matrix with two texture zeros and an equality between elements impose constraints on $M_{\nu}$ and reduce the degree of freedom by six. Therefore, these textures are more predictive as compared to the models with two texture zeros, two vanishing cofactors and hybrid textures.
\item On the experimental side, the neutrino mass ordering and the absolute neutrino mass scale are unknown. 
Since, we are expected to clearly distinguish the neutrino mass spectrum in the near future, a detailed phenomenological investigation of these textures is expected to be quite rewarding as these textures are able to distinguish between normal ordering (NO) and inverted ordering (IO) of neutrino mass spectrum.  
\end{itemize}

The plan of this paper is as follows: Section II describes the formalism and outlines the numerical analysis. Section III presents the numerical results. Section IV is devoted to the realization of a viable texture in the framework of type-II seesaw mechanism using $S_3 \times Z_3$ symmetry. The main conclusions are summarized in Section V.

\section{Formalism and Analysis}
For Majorana neutrinos, the neutrino mass matrix is complex symmetric with six independent elements. There are $^2C_6=15$ possible textures for $M_{\nu}$ with two texture zeros out of which only seven are phenomenologically compatible \cite{texture1} with the current neutrino oscillation data at $3\sigma$ C.L. These seven two zero textures in combination with one equality between two nonzero elements lead to a total of forty-two textures for $M_{\nu}$ tabulated in Table \ref{table:equal}. These forty-two textures have been classified into seven different classes depending on the position of the texture zeros in $M_{\nu}$ and each class comprises six different textures. 
\begin{table}[h]
\caption{Forty-two possible texture structures for $M_{\nu}$ having two texture zeros and one equality between two nonzero elements where $\Delta$ denote two equal nonzero elements and $\times$ represents a nonzero arbitrary entry.}
\begin{scriptsize}
\begin{tabular}{ccccccc}
 \hline \hline
Class & I & II & III & IV & V & VI \\
 \hline
$A_{1}$ & $\left(\begin{array}{ccc}
 0 & 0 & \Delta \\
 0 & \times & \Delta \\
 \Delta & \Delta & \times \\
\end{array}
\right)$ & $\left(\begin{array}{ccc}
 0 & 0 & \Delta \\
 0 & \Delta & \times \\
 \Delta & \times & \times \\
\end{array}
\right)$ & $\left(\begin{array}{ccc}
 0 & 0 & \Delta \\
 0 & \times & \times \\
 \Delta & \times & \Delta \\
\end{array}
\right)$ & $\left(\begin{array}{ccc}
 0 & 0 & \times \\
 0 & \Delta & \Delta \\
 \times & \Delta & \times \\
\end{array}
\right)$ & $\left(\begin{array}{ccc}
 0 & 0 & \times \\
 0 & \Delta & \times \\
 \times & \times & \Delta \\
\end{array}
\right)$ & $\left(\begin{array}{ccc}
 0 & 0 & \times \\
 0 & \times & \Delta \\
 \times & \Delta & \Delta \\
\end{array}
\right)$\\
$A_{2}$ & $\left(\begin{array}{ccc}
 0 & \Delta & 0 \\
 \Delta & \Delta & \times \\
 0 & \times & \times \\
\end{array}
\right)$ & $\left(\begin{array}{ccc}
 0 & \Delta & 0 \\
 \Delta & \times & \Delta \\
 0 & \Delta & \times \\
\end{array}
\right)$ & $\left(\begin{array}{ccc}
 0 & \Delta & 0 \\
 \Delta & \times & \times \\
 0 & \times & \Delta \\
\end{array}
\right)$ & $\left(\begin{array}{ccc}
 0 & \times & 0 \\
 \times & \Delta & \Delta \\
 0 & \Delta & \times \\
\end{array}
\right)$ & $\left(\begin{array}{ccc}
 0 & \times & 0 \\
 \times & \Delta & \times \\
 0 & \times & \Delta \\
\end{array}
\right)$ & $\left(\begin{array}{ccc}
 0 & \times & 0 \\
 \times & \times & \Delta \\
 0 & \Delta & \Delta \\
\end{array}
\right)$\\
$B_{1}$ & $\left(\begin{array}{ccc}
 \Delta & \Delta & 0 \\
 \Delta & 0 & \times \\
 0 & \times & \times \\
\end{array}
\right)$ & $\left(\begin{array}{ccc}
 \Delta & \times & 0 \\
 \times & 0 & \Delta \\
 0 & \Delta & \times \\
\end{array}
\right)$ & $\left(\begin{array}{ccc}
 \Delta & \times & 0 \\
 \times & 0 & \times \\
 0 & \times & \Delta \\
\end{array}
\right)$ & $\left(\begin{array}{ccc}
 \times & \Delta & 0 \\
 \Delta & 0 & \Delta \\
 0 & \Delta & \times \\
\end{array}
\right)$ & $\left(\begin{array}{ccc}
 \times & \Delta & 0 \\
 \Delta & 0 & \times \\
 0 & \times & \Delta \\
\end{array}
\right)$ & $\left(\begin{array}{ccc}
 \times & \times & 0 \\
 \times & 0 & \Delta \\
 0 & \Delta & \Delta \\
\end{array}
\right)$\\
$B_{2}$ : & $\left(\begin{array}{ccc}
 \Delta & 0 & \Delta \\
 0 & \times & \times \\
 \Delta & \times & 0 \\
\end{array}
\right)$ & $\left(\begin{array}{ccc}
  \Delta & 0 & \times \\
 0 & \Delta & \times \\
 \times & \times & 0 \\
\end{array}
\right)$ & $\left(\begin{array}{ccc}
 \Delta & 0 & \times \\
 0 & \times & \Delta \\
 \times & \Delta & 0 \\
\end{array}
\right)$ & $\left(\begin{array}{ccc}
 \times & 0 & \Delta \\
 0 & \Delta & \times \\
 \Delta & \times & 0 \\
\end{array}
\right)$ & $\left(\begin{array}{ccc}
 \times & 0 & \Delta \\
 0 & \times & \Delta \\
 \Delta & \Delta & 0 \\
\end{array}
\right)$ & $\left(\begin{array}{ccc}
 \times & 0 & \times \\
 0 & \Delta & \Delta \\
 \times & \Delta & 0 \\
\end{array}
\right)$\\
$B_{3}$ & $\left(\begin{array}{ccc}
 \Delta & 0 & \Delta \\
 0 & 0 & \times \\
 \Delta & \times & \times \\
\end{array}
\right)$ & $\left(\begin{array}{ccc}
 \Delta & 0 & \times \\
 0 & 0 & \Delta \\
 \times & \Delta & \times \\
\end{array}
\right)$ & $\left(\begin{array}{ccc}
 \Delta & 0 & \times \\
 0 & 0 & \times \\
 \times & \times & \Delta \\
\end{array}
\right)$ & $\left(\begin{array}{ccc}
 \times & 0 & \Delta \\
 0 & 0 & \Delta \\
 \Delta & \Delta & \times \\
\end{array}
\right)$ & $\left(\begin{array}{ccc}
 \times & 0 & \Delta \\
 0 & 0 & \times \\
 \Delta & \times & \Delta \\
\end{array}
\right)$ & $\left(\begin{array}{ccc}
 \times & 0 & \times \\
 0 & 0 & \Delta \\
 \times & \Delta & \Delta \\
\end{array}
\right)$\\
$B_{4}$ & $\left(\begin{array}{ccc}
 \Delta & \Delta & 0 \\
 \Delta & \times & \times \\
 0 & \times & 0 \\
\end{array}
\right)$ & $\left(\begin{array}{ccc}
 \Delta & \times & 0 \\
 \times & \Delta & \times \\
 0 & \times & 0 \\
\end{array}
\right)$ & $\left(\begin{array}{ccc}
  \Delta & \times & 0 \\
 \times & \times & \Delta \\
 0 & \Delta & 0 \\
\end{array}
\right)$ & $\left(\begin{array}{ccc}
  \times & \Delta & 0 \\
 \Delta & \Delta & \times \\
 0 & \times & 0 \\
\end{array}
\right)$ & $\left(\begin{array}{ccc}
  \times & \Delta & 0 \\
 \Delta & \times & \Delta \\
 0 & \Delta & 0 \\
\end{array}
\right)$ & $\left(\begin{array}{ccc}
 \times & \times & 0 \\
 \times & \Delta & \Delta \\
 0 & \Delta & 0 \\
\end{array}
\right)$\\
$C $ & $\left(\begin{array}{ccc}
 \Delta & \Delta & \times \\
 \Delta & 0 & \times \\
 \times & \times & 0 \\
\end{array}
\right)$ & $\left(\begin{array}{ccc}
 \Delta & \times & \Delta \\
 \times & 0 & \times \\
 \Delta & \times & 0 \\
\end{array}
\right)$ & $\left(\begin{array}{ccc}
 \Delta & \times & \times \\
 \times & 0 & \Delta \\
 \times & \Delta & 0 \\
\end{array}
\right)$ & $\left(\begin{array}{ccc}
 \times & \Delta & \Delta \\
 \Delta & 0 & \times \\
 \Delta & \times & 0 \\
\end{array}
\right)$ & $\left(\begin{array}{ccc}
 \times & \Delta & \times \\
 \Delta & 0 & \Delta \\
 \times & \Delta & 0 \\
\end{array}
\right)$ & $\left(\begin{array}{ccc}
\times & \times & \Delta \\
 \times & 0 & \Delta \\
 \Delta & \Delta & 0 \\
\end{array}
\right)$\\
 \hline
 \end{tabular}
\label{table:equal}
\end{scriptsize}
\end{table}
   
In the flavor basis, where the charged lepton mass matrix $M_{l}$ is diagonal, the complex symmetric Majorana neutrino mass matrix $M_{\nu}$ can be diagonalized by an unitary matrix:
\begin{equation}
M_{\nu}=V'M_{\nu}^{diag}V'^{T}
\end{equation}
where $M_{\nu}^{diag}=\textrm{Diag}(m_1,m_2,m_3)$ and $V'$ is the unitary matrix. The unitary matrix $V'$ can be parametrized as
\begin{equation}
V'=P_{l} V ~~\textrm{with}~~ V=U P_{\nu}
\end{equation}
where
\begin{align}
U&=\left(
\begin{array}{ccc}
 c_{12} c_{13} & c_{13} s_{12} & e^{-i \delta } s_{13} \\
 -c_{23} s_{12}-e^{i \delta } c_{12} s_{13} s_{23} & c_{12} c_{23}-e^{i \delta } s_{12} s_{13} s_{23}
   & c_{13} s_{23} \\
 s_{12} s_{23}-e^{i \delta } c_{12} c_{23} s_{13} & -e^{i \delta }
  c_{23} s_{12} s_{13}-c_{12} s_{23} & c_{13} c_{23}
\end{array}
\right),\\
P_{\nu}&=\left(
\begin{array}{ccc}
1 & 0 & 0 \\
0 & e^{i \alpha}&0 \\
0 & 0 & e^{i \beta}
\end{array}
\right),~~ P_{l}=
\left(
\begin{array}{ccc}
e^{i \phi_{e}} & 0 & 0 \\
0 & e^{i \phi_{\mu}}&0 \\
0 & 0 & e^{i \phi_{\tau}}
\end{array}
\right)
\end{align}
with $c_{ij}=\cos \theta_{ij}$, $s_{ij}=\sin \theta_{ij}$ and  $\delta$ is the Dirac-type CP-violating phase. The phase matrix $P_{\nu}$ contains two Majorana-type CP-violating phases $\alpha$ and $\beta$ while the phase matrix $P_l$ is physically unobservable. The neutrino mixing matrix $V$ is also known as the Pontecorvo-Maki-Nakagawa-Sakata (PMNS) \cite{pmns} matrix. The Majorana neutrino mass matrix, containing twelve neutrino parameters viz. three mass eigenvalues, three mixing angles, three CP-violating phases and three unphysical phases, can be written as
\begin{equation}
M_{\nu}=P_{l}U P_{\nu}M_{\nu}^{diag}P_{\nu}^{T}U^{T}P_{l}^{T}.
\end{equation}
The Dirac CP-violation in neutrino oscillation experiments can be expressed in terms of Jarlskog rephasing invariant quantity $J_{CP}$ \cite{jcp} with
\begin{equation}
J_{CP}=\textrm{Im}\lbrace U_{11} U_{22} U^{*}_{12} U^{*}_{21}\rbrace=\sin\theta_{12} \sin\theta_{23} \sin\theta_{13} \cos\theta_{12} \cos\theta_{23} \cos^{2}\theta_{13} \sin\delta .
\end{equation}
The parameter which determines the rate of neutrinoless double beta decay, known as effective Majorana neutrino mass $|M_{ee}|$,  is given by
\begin{equation}
|M_{ee}|=|m_1 U^{2}_{e1}+m_2 U^{2}_{e2}+m_3 U^{2}_{e3}|.
\end{equation}
There are many experiments such as CUORICINO \cite{cuori}, CUORE \cite{cuore}, MAJORANA \cite{majorana}, SuperNEMO \cite{nemo}, EXO \cite{exo} with a targeted sensitivity upto $0.01$ eV for observing $|M_{ee}|$.\\
Further, cosmological observations provide more stringent constraints on absolute neutrino mass scale. Recent Planck data \cite{planck} combined with baryon acoustic oscillation (BAO) measurements provide a stringent constraint on the sum of neutrino masses $\sum m_{i}\leq 0.12$ eV at 95$\%$ confidence level (C.L.).

There exists a $\mu$-$\tau$ permutation symmetry between different textures of neutrino mass matrices and the corresponding permutation matrix has the following form:
\begin{equation}
P_{23}=\left(
\begin{array}{ccc}
 1 & 0 & 0 \\
 0 & 0 & 1 \\
 0 & 1 & 0 \\
\end{array}
\right).
\end{equation}   
The hybrid neutrino mass matrices are related to each other by the transformation
\begin{equation}
M^{'}_{\nu}=P_{23} M_{\nu} P^{T}_{23}
\end{equation}
leading to the following relations between the neutrino oscillation parameters:
\begin{equation}
\theta^{'}_{12}=\theta_{12},~~ \theta^{'}_{13}=\theta_{13},~~ \theta^{'}_{23}=\frac{\pi}{2}-\theta_{23},~~ \delta^{'}=\pi-\delta.
\end{equation}
Neutrino mass matrices with two texture zeros and two equal elements possessing $\mu$-$\tau$ permutation symmetry are
\begin{align}
& A_1^{I}\leftrightarrow A_2^{II},~ A_1^{II}\leftrightarrow A_2^{III}, ~A_1^{III}\leftrightarrow A_2^{I},~A_1^{IV}\leftrightarrow A_2^{VI},~ A_1^{V}\leftrightarrow A_2^{V}, ~A_1^{VI}\leftrightarrow A_2^{IV},\nonumber\\
& B_1^{I}\leftrightarrow B_2^{I},~ B_1^{II}\leftrightarrow B_2^{III}, ~B_1^{III}\leftrightarrow B_2^{II},~B_1^{IV}\leftrightarrow B_2^{V},~ B_1^{V}\leftrightarrow B_2^{IV}, ~B_1^{VI}\leftrightarrow B_2^{VI},\\
& B_3^{I}\leftrightarrow B_4^{I},~ B_3^{II}\leftrightarrow B_4^{III}, ~B_3^{III}\leftrightarrow B_4^{II},~B_3^{IV}\leftrightarrow B_4^{V},~ B_3^{V}\leftrightarrow B_4^{IV}, ~B_3^{VI}\leftrightarrow B_4^{VI},\nonumber\\
& C^{I}\leftrightarrow C^{II},~ C^{III}\leftrightarrow C^{III}, ~C^{IV}\leftrightarrow C^{IV},~C^{V}\leftrightarrow C^{VI}.\nonumber
\end{align}
Note that textures $C^{III}$ and $C^{IV}$ transform onto themselves under the $\mu$-$\tau$ permutation. Thus, out of a total forty-two textures, only twenty-two are independent.
\begin{table}[h]
\caption{Current Neutrino oscillation parameters from global fits \cite{data} with $\Delta m^{2}_{3 l}\equiv \Delta m^{2}_{31}>0$ for NO and $\Delta m^{2}_{3 l}\equiv \Delta m^{2}_{32}=-\Delta m^{2}_{23}<0$ for IO.}
\begin{center}
\begin{tabular}{lll}
 \hline \hline
Neutrino Parameter & Normal Ordering (best fit)& Inverted Ordering ($\Delta \chi^2=4.7$)\\
  & bfp $\pm 1 \sigma$ ~~~~~~~~~ $3\sigma$ range  & bfp $\pm 1 \sigma$~~~~~~~~~~ $3\sigma$ range \\
 \hline
$\theta_{12}^{\circ}$ & $33.82^{+0.78}_{-0.76}$ ~~~~~~ $31.61 \rightarrow 36.27$ & $33.82^{+0.78}_{-0.76}$ ~~~~~~~ $31.61 \rightarrow 36.27$ \\
$\theta_{23}^{\circ}$ & $49.6^{+1.0}_{-1.2}$  ~~~~~~~~~~~$40.3 \rightarrow 52.4$ & $49.8^{+1.0}_{-1.1}$ ~~~~~~~~~~$40.6 \rightarrow 52.5$ \\
$\theta_{13}^{\circ}$ & $8.61^{+0.13}_{-0.13}$ ~~~~~ ~~~~$8.22 \rightarrow 8.99$ & $8.65^{+0.13}_{-0.13}$ ~~~~~~~~ $8.27 \rightarrow 9.03$ \\
$\delta_{CP}^{\circ}$ & $215^{+40}_{-29}$ ~~~~~~~~~~~~ $125 \rightarrow 392$ & $284^{+ 27}_{-29}$ ~~~~~~~~~~~ $196 \rightarrow 360$ \\
$\Delta m^{2}_{21}/10^{-5} eV^2 $ & $7.39^{+0.21}_{-0.20}$ ~~~~~~~~~~$6.79 \rightarrow 8.01$ & $7.39^{+0.21}_{-0.20}$ ~~~~~~~~ $6.79 \rightarrow 8.01$ \\
$\Delta m^{2}_{3 l}/10^{-3} eV^2 $ & $+2.525^{+0.033}_{-0.032}$ ~ $
+2.427 \rightarrow +2.625$ & $-2.512^{+0.034}_{-0.032}$~  $-2.611 \rightarrow -2.412$ \\
 \hline 
 \end{tabular}
\label{table:dat}
\end{center}
\end{table}

\subsection{Two Texture Zeros and an Equality Between the Elements of $M_{\nu}$}
The simultaneous existence of two texture zeros at the $(a,b), (c,d)$-positions and an equality between the elements at the $(p,q), (r,s)$-positions of neutrino mass matrix implies the following constraints:
\begin{equation}
M_{\nu (a,b)}=0,~~M_{\nu (c,d)}=0~~ \textrm{and} ~~M_{\nu (p,q)}-M_{\nu (r,s)}=0.
\end{equation}
The above equations can be rewritten as
\begin{eqnarray}
m_1 A_1+m_2 A_2 e^{2i\alpha}+m_3 A_3 e^{2i\beta}&=&0,\\
m_1 B_1+m_2 B_2 e^{2i\alpha}+m_3 B_3 e^{2i\beta}&=&0,\\
m_1 C_1+m_2 C_2 e^{2i\alpha}+m_3 C_3 e^{2i\beta}&=&0,
\end{eqnarray}
where $A_i=U_{ai} U_{bi},B_i=U_{ci} U_{di}$ and $C_i=Q U_{pi} U_{qi}-U_{ri} U_{si}$ with $Q=e^{\phi_p+\phi_q-(\phi_r+\phi_s)}$.
By solving the texture zero conditions given in Eqs. (13) and (14) for two mass ratios, we obtain
\begin{eqnarray}
\frac{m_2}{m_1}e^{2i\alpha}=\frac{A_3 B_1-A_1 B_3}{A_2 B_3-A_3 B_2} ~~\textrm{and}~~ \frac{m_3}{m_1}e^{2i\beta}=\frac{A_2 B_1-A_1 B_2}{A_3 B_2-A_2 B_3}.
\end{eqnarray}
The magnitudes of two mass ratios are given by
\begin{eqnarray}
\zeta\equiv\left|\frac{m_2}{m_1}\right|=\left| \frac{A_3 B_1-A_1 B_3}{A_2 B_3-A_3 B_2} \right|~~
\textrm{and}~~ \xi\equiv\left|\frac{m_3}{m_1}\right|=\left| \frac{A_2 B_1-A_1 B_2}{A_3 B_2-A_2 B_3} \right|
\end{eqnarray}
while the CP-violating Majorana phases $\alpha$ and $\beta$ are
\begin{eqnarray}
\alpha=\frac{1}{2}arg\left(\frac{A_3 B_1-A_1 B_3}{A_2 B_3-A_3 B_2} \right)~~\textrm{and}~~
\beta=\frac{1}{2}arg\left(\frac{A_2 B_1-A_1 B_2}{A_3 B_2-A_2 B_3} \right),
\end{eqnarray}  
respectively.
The neutrino mass eigenvalues $m_2$ and $m_3$ can be calculated form the mass-squared differences $\Delta m_{21}^2$ and $\Delta m_{31}^2$ $(\Delta m_{23}^2)$ for NO (IO) using the following relations:
\begin{eqnarray}
m_2&=&\sqrt{m_1^2+\Delta m_{21}^2}, ~~~m_3=\sqrt{m_1^2+\Delta m_{31}^2} ~~~~~~~~~~~~~~~~~~~~~~~~~~~\textrm{for NO}, \nonumber \\
\textrm{and}~~~~ m_2&=&\sqrt{m_1^2+\Delta m_{21}^2}, ~~~m_3=\sqrt{m_1^2+\Delta m_{21}^2-\Delta m_{23}^2} ~~~~~~~~~~~~~~~~\textrm{for IO}
\end{eqnarray} 
where $\Delta m_{ij}^2=m^2_i-m^2_j$, $m_1<m_2<m_3$ for NO and $m_3<m_1<m_2$ for IO.
By using mass ratios given in Eq. (17), the ratios of mass-squared differences ($R_{\nu}$) are given by
\begin{equation}
R_{\nu}\equiv \frac{\Delta m_{21}^2}{\Delta m_{31}^2}=\frac{\zeta^2-1}{\xi^2-1}~~~~ \textrm{and~~ } R_{\nu}\equiv \frac{\Delta m_{21}^2}{\Delta m_{23}^2}=\frac{\zeta^2-1}{\zeta^2-\xi^2}
\end{equation}
for NO and IO, respectively. Since, $\Delta m_{21}^2$ and $\Delta m_{31}^2$ $(\Delta m_{32}^2)$ for NO (IO) are experimentally available, the parameter $R_{\nu}$ should lie within its experimentally allowed range for a texture to be compatible with the current neutrino oscillation data tabulated in Table II.\\
The texture equality constraint can be imposed by substituting Eq.(16) in Eq.(15) which yields the following non-trivial condition:
\begin{equation}
\Lambda \equiv A_1 B_3 C_2+A_2 B_1 C_3+A_3 B_2 C_1-A_1 B_2 C_3-A_2 B_3 C_1-A_3 B_1 C_2=0.
\end{equation} 
The above complex equation leads to the following two equations:
\begin{equation}
Re(\Lambda)=0~~ ~\textrm{and}~~~ Im(\Lambda)=0
\end{equation}
which can be further solved to calculate the mixing angle $\theta_{12}$ and one phase angle by using these constraints for each texture.\\
As a result it is possible to calculate the six parameters viz. the two mass ratios, the two Majorana phases, $\theta_{12}$ mixing angle and one phase angle in terms of other parameters. The analysis of each viable texture is described in the following:
\begin{center}
\subsubsection{\bf{Class $A_{1}$}}
\end{center}
For class $A_1$, there are two texture zeros corresponding to elements $M_{ee}$, $M_{e\mu}$ of neutrino mass matrix and solving these, by using Eq. (16), leads to the following mass ratios:
\begin{eqnarray}
\frac{m_2}{m_1} e^{2 i \alpha }&=&\frac{\cos\theta _{12} \left(\sin\theta _{13} \cos
   \theta _{23}+e^{i \delta } \sin \theta _{23} \cot\theta
   _{12}\right)}{\sin \theta _{13} \cos \theta _{12} \cos \theta _{23}-e^{i \delta } \sin \theta _{12} \sin \theta _{23}}, \\
\frac{m_3}{m_1} e^{2 i \beta }&=&\frac{ \cos\theta _{12} \cos\theta _{13} \cos \theta _{23} \cot\theta _{13}}{-\sin\theta _{13} \cos
 \theta _{12} \cos \theta _{23}+e^{i \delta } \sin\theta _{12} \sin \theta _{23}} e^{2 i \delta}.
\end{eqnarray}
There are three viable textures namely $A_{1}^{IV}, A_{1}^{V}, A_{1}^{VI}$ for class $A_1$ and their predictions have been described  below.\\  
\textbf{Texture $A_{1}^{IV}$}: The texture structure $A_{1}^{IV}$  has one equality at the (2,2), (2,3)-positions i.e. $M_{\mu \mu}=M_{\mu \tau}$ along with texture zeros at the (1,1), (1,2)-positions and can be represented as
\begin{eqnarray}
A_{1}^{IV}=\left(\begin{array}{ccc}
 0 & 0 & \times \\
 0 & \Delta & \Delta \\
 \times & \Delta & \times \\
\end{array}
\right)~.
\end{eqnarray}
 Solving, Eq. (22) for  $M_{\mu \mu}=M_{\mu \tau}$ yields the following relations:
\begin{equation}
\sin 2\theta_{12}=-\frac{P}{\sqrt{P^2+Q^2}}
~~ \textrm{and}~~
\sin\delta=\frac{U}{\sqrt{U^2+V^2}}
\end{equation}

where
\begin{align}
P&=8 \sin2 \theta _{23} \left(2 \cos\theta _{23} \sin(\delta +\phi _{\mu})+\cos(\delta -\theta _{23}+\phi _{\tau })-\cos(\delta +\theta _{23}+\phi_{\tau })\right),\nonumber\\
Q&=\csc\theta _{13}(4 \cos\theta _{23} (-\sin2 \delta \cos\phi _{\mu }-\cos2\delta \sin\phi _{\mu }\nonumber\\
&+\sin^2\theta_{13}(4 \sin \theta _{23} \cos\theta _{23} \sin\phi_{\tau }+3 \sin \phi _{\mu })) \nonumber\\
& -\cos3 \theta _{23} (-4\sin (2 \delta +\phi _{\mu })-\sin (2 \theta _{13}-\phi _{\mu })+\sin(2\theta _{13}+\phi _{\mu })-2 \sin\phi _{\mu })\nonumber \\
&+8 \sin \theta_{23}(\cos 2 \theta _{13}+\cos 2 \theta _{23}) \sin(2\delta +\phi _{\tau })),\\
U&=\sin 2 \theta _{23} \cos ^2\theta _{13} \sin(\phi _{\mu }-\phi _{\tau}),\nonumber\\
V&=\frac{1}{2} \left(\sin 4 \theta _{23} \cos ^2\theta _{13} \cos(\phi _{\mu}-\phi _{\tau })-4 \sin ^2\theta _{13} \left(\sin 2 \theta _{23} \cos (\phi _{\mu }-\phi _{\tau })+1\right)\right).\nonumber
\end{align}

\textbf{Texture $A_{1}^{V}$}: Texture $A_{1}^{V}$, with two equal elements at the (2,2), (3,3)-positions and two texture zeros at the (1,1), (1,2)-positions, can be represented as
\begin{eqnarray}
A_{1}^{V}= \left(\begin{array}{ccc}
 0 & 0 & \times \\
 0 & \Delta & \times \\
 \times & \times & \Delta \\
\end{array}
\right).
\end{eqnarray}
Using the $M_{\mu \mu}=M_{\tau \tau}$ condition (Eq. (22)) yields
\begin{equation}
\sin 2\theta_{12}=-\frac{P}{\sqrt{P^2+Q^2}}~~ \textrm{and}~~ \sin\delta=-\frac{U}{\sqrt{U^2+V^2}}
\end{equation}
where
\begin{align}
P&=4 (\sin\theta _{23}\left(\cos(\delta +2 \phi _{\mu })-3 \cos(\delta
   +2 \phi _{\tau })\right)\nonumber\\
   &+2 \sin 3 \theta _{23} \cos(\phi _{\mu }-\phi _{\tau
   }) \cos (\delta +\phi _{\mu }+\phi _{\tau })),\nonumber \\
Q&=\sin\theta _{13} \cos\theta _{23}(-8 \sin^2\theta _{23}(\csc^2\theta _{13} \cos(2(\delta +\phi _{\mu }))+\cos
  2 \phi _{\tau })\nonumber\\
  &+\frac{1}{2}(5 \cos2 \theta _{13}+8 \cos2\theta _{23}+3) \csc^2\theta _{13} \cos(2 (\delta +\phi _{\tau
   }))\nonumber \\
   & +11 \sin2\delta \sin2\phi _{\tau }- 11 \cos 2 \delta  \cos 2 \phi _{\tau }+8 \cos ^2\theta _{23} \cos2 \phi _{\mu }),\\
U&=8 (2 \cos 2 \theta _{23} \cos ^2\theta _{13}+\cos2 \theta_{13}-1) \sin2(\phi _{\mu }-\phi _{\tau },\nonumber) \\
V&=2 (-6 \cos^2\theta _{13} \cos2(\phi _{\mu }-\phi _{\tau })-4(\cos ^2\theta _{13} \cos4\theta _{23}-4 \sin ^2\theta_{13} \cos 2 \theta _{23})\nonumber\\
& ~~ \cos ^2(\phi _{\mu }-\phi _{\tau })+\cos2 \theta _{13}+1).\nonumber  
\end{align}

\textbf{Texture $A_{1}^{VI}$}: Texture $A_{1}^{VI}$, having one equality at the (2,3), (3,3)-positions along with two texture zeros at the (1,1), (1,2)-positions can be represented as
\begin{eqnarray}
A_{1}^{VI}= \left(\begin{array}{ccc}
 0 & 0 & \times \\
 0 & \times & \Delta \\
 \times & \Delta & \Delta \\
\end{array}
\right).
\end{eqnarray}
Solving Eq. (22) for texture $A_{1}^{VI}$ yields
\begin{small}
\begin{equation}
\sin 2\theta_{12}=-\frac{P}{\sqrt{P^2+Q^2}}~~\textrm{and}~~\sin\delta=\frac{\sin \left(\phi _{\mu }-\phi _{\tau }\right)}{\sqrt{\cos ^2 2 \theta _{23} \cos
   ^2\left(\phi _{\mu }-\phi _{\tau }\right)+\sin ^2\left(\phi _{\mu }-\phi _{\tau }\right)}}
\end{equation}
\end{small}
where
\begin{align}
P&= 4 \sin ^2\theta _{13} \sin ^2\theta _{23} (\cos \theta _{23} \sin(\delta +\phi _{\mu })+\sin \theta _{23} \sin(\delta +\phi _{\tau})),\nonumber \\
Q&=\sin \theta _{13}(\sin \theta _{23} ((\cos2 \theta_{13}+\cos2 \theta _{23}) \sin(2 \delta +\phi _{\mu })+\sin2 \theta _{23} \sin ^2\theta _{13} \sin\phi _{\tau })\\
&-\cos\theta _{23}(2 \cos2 \theta _{13}+\cos2 \theta_{23}-1) \sin(2 \delta +\phi _{\tau })+2 \sin \theta _{23} \sin^2\theta _{13} \cos ^2\theta _{23} \sin\phi _{\mu }).\nonumber
\end{align}

\subsubsection{\bf{Class $A_{2}$}}

For class $A_2$ each texture has two texture zeros at the (1,1) and (1,3)-positions.  Using Eq. (16) for $M_{ee}=0, M_{e\tau}=0$ gives the following mass ratios:
\begin{eqnarray}
\frac{m_2}{m_1} e^{2 i \alpha }&=& \frac{ \cos\theta _{12} \left(\sin\theta _{13} \sin\theta _{23}-e^{i \delta } \cos\theta _{23} \cot\theta
   _{12}\right)}{\sin\theta _{13} \sin \theta _{23} \cos\theta _{12}+e^{i \delta } \sin \theta _{12} \cos\theta _{23}},\\
\frac{m_3}{m_1} e^{2 i \beta }&=& -\frac{ \sin\theta _{23} \cos \theta _{12} \cos\theta _{13} \cot\theta _{13}}{\sin\theta _{13} \sin
  \theta _{23} \cos \theta _{12}+e^{i \delta } \sin \theta _{12}
   \cos \theta _{23}} e^{2 i \delta }.
\end{eqnarray}
Only three textures of class $A_{2}$  are viable and their predictions are described in the following:\\
\textbf{Texture $A_{2}^{IV}$}: Texture $A_{2}^{IV}$, having two equal elements at the (2,2), (2,3)-positions and two zeros at the (1,1), (1,3)-positions, can be represented by 
\begin{eqnarray}
A_{2}^{IV}= \left(\begin{array}{ccc}
 0 & \times & 0 \\
 \times & \Delta & \Delta \\
 0 & \Delta & \times \\
\end{array}
\right).
\end{eqnarray}
Use of the equality constraint (Eq. (22)) for texture $A_{2}^{IV}$ yields the following relations:
\begin{small}
\begin{equation}
\sin 2\theta_{12}=\frac{P}{\sqrt{P^2+Q^2}}~~\textrm{and}~~\sin\delta=\frac{\sin \left(\phi _{\mu }-\phi _{\tau }\right)}{\sqrt{\cos ^22 \theta _{23} \cos
   ^2\left(\phi _{\mu }-\phi _{\tau }\right)+\sin ^2\left(\phi _{\mu }-\phi _{\tau }\right)}}
\end{equation}
\end{small}
where
\begin{align}
P&= \frac{1}{2} \cos ^2\theta _{23} (2 \sin\theta _{23} \cos (\delta +\phi _{\mu }+\phi _{\tau })+\cos(\delta -\theta _{23}+2 \phi _{\mu })+\cos(\delta+\theta _{23}+2 \phi _{\mu })),\nonumber\\
Q&= \csc\theta _{13}(\cos\theta _{23}(\cos2 \delta (\cos2 \theta _{13}-\cos2 \theta _{23})+2 \sin ^2\theta _{13}
   \sin ^2\theta _{23}) \cos(\phi _{\mu }+\phi _{\tau }) \nonumber\\
   &+\sin2 \delta (\cos\theta _{23}(\cos2 \theta _{23}-\cos 2 \theta
   _{13}) \sin (\phi _{\mu }+\phi _{\tau })+2 \sin\theta _{23} \cos2 \theta _{13} \sin 2 \phi _{\mu })\\
   &+2 \sin\theta _{23}(\cos2 \phi _{\mu } (\cos ^2 \theta _{23} (\cos (2 \delta
   )+\sin ^2 \theta _{13})-\cos (2 \delta ) \cos 2 \theta _{13})\nonumber\\
   & -4 \sin\delta \cos\delta \cos ^2\theta _{23} \sin \phi _{\mu } \cos
   \phi _{\mu })).\nonumber
\end{align}

\textbf{Texture $A_{2}^{V}$}: The neutrino mass matrix corresponding to texture $A_{2}^{V}$ is given by
\begin{eqnarray}
A_{2}^{V}= \left(\begin{array}{ccc}
 0 & \times & 0 \\
 \times & \Delta & \times \\
 0 & \times & \Delta \\
\end{array}
\right).
\end{eqnarray}
The imposition of texture equality constraint$M_{\mu \mu}=M_{\tau \tau}$ (Eq. (22)) provides the following relations:
\begin{equation}
\sin 2\theta_{12}=-\frac{P}{\sqrt{P^2+Q^2}}~~\textrm{and}~~\sin\delta=\frac{U}{\sqrt{U^2+V^2}}
\end{equation}
where
\begin{align}
P&=2 \cos\theta _{23} \left(\sin ^2\theta _{23} \cos \left(\delta +2 \phi _{\tau
   }\right)-\cos ^2\theta _{23} \cos \left(\delta +2 \phi _{\mu }\right)\right),\nonumber\\
Q&=\sin\theta _{13} \sin\theta _{23}(\csc ^2\theta _{13}
   ((\cos ^2\theta _{23}-\cos2 \theta _{13}) \cos2
   (\delta +\phi _{\mu })\nonumber\\
   &+\cos ^2\theta _{23} \cos2(\delta+\phi _{\tau }))+\cos ^2\theta _{23} \cos2 \phi _{\mu
   } -\sin ^2\theta _{23} \cos 2 \phi _{\tau }),\nonumber\\ 
U&=\left(\cos2 \theta _{23} \cot ^2\theta _{13}+1\right) \sin2 \left(\phi
   _{\mu }-\phi _{\tau }\right), \\
V&=\frac{1}{4}(-2 (\cos4 \theta _{23}+2(\cos2 \theta _{13}-2 \cos
   ^2\theta _{23}) \cos 2 \theta _{23} \csc ^2\theta _{13}) \cos ^2(\phi _{\mu }-\phi _{\tau })\nonumber\\
   &-4 \cos2 \theta _{13}
   \csc ^2\theta _{13} \sin ^2(\phi _{\mu }-\phi _{\tau }) +\cos2(\phi_{\mu }-\phi _{\tau })-3).\nonumber
\end{align}

\textbf{Texture $A_{2}^{VI}$}: For texture $A_{2}^{VI}$, the neutrino mass matrix is given by
\begin{eqnarray}
A_{2}^{VI}= \left(\begin{array}{ccc}
 0 & \times & 0 \\
 \times & \times & \Delta \\
 0 & \Delta & \Delta \\
\end{array}
\right).
\end{eqnarray}
The mixing angles and the Dirac-type CP-violating phase are given by
\begin{equation}
\sin 2\theta_{12}=\frac{P}{\sqrt{P^2+Q^2}}~~\textrm{and}~~\sin\delta=\frac{U}{\sqrt{U^2+V^2}}
\end{equation}
where
\begin{align}
P&=\sin\theta _{13} \sin 2 \theta _{23} \left(\cos \left(\delta -\theta _{23}+\phi
   _{\mu }\right)+\cos \left(\delta +\theta _{23}+\phi _{\mu }\right)+2 \sin\theta _{23} \cos
   \left(\delta +\phi _{\tau }\right)\right),\nonumber \\
Q&=\sin 2 \delta \cos \theta _{23} \left(\left(\cos 2 \theta _{23}-\cos 2\theta _{13}\right) \sin\phi _{\mu }+\sin 2 \theta _{23} \sin\phi _{\tau }\right)\nonumber\\
&+\cos\theta _{23} \cos \phi _{\mu }
   (\cos 2 \delta (\cos 2 \theta _{13}-\cos 2 \theta _{23})+2 \sin ^2\theta _{13} \sin ^2\theta _{23})\nonumber\\
   &+\cos \phi _{\tau }
   \left(2 \sin ^2\theta _{13} \sin ^3\theta _{23}-2 \cos 2 \delta \sin
   \theta _{23} \cos ^2\theta _{23}\right),\nonumber\\
U&= \sin 2 \theta _{23} \cos ^2\theta _{13} \sin \left(\phi _{\mu }-\phi _{\tau
   }\right),\\
V&= \frac{1}{4} \left(2 \left(4 \sin ^2\theta _{13} \sin 2 \theta _{23}+\sin 4
   \theta _{23} \cos ^2\theta _{13}\right) \cos \left(\phi _{\mu }-\phi _{\tau
   }\right)+8 \sin ^2\theta _{13}\right).\nonumber
\end{align}

\begin{center}
\subsubsection{\bf{Class $C$}}
\end{center}
All textures of class $C$ have two texture zeros at the (2,2) and (3,3)-positions and Eq. (16) can be used to calculate the two mass ratios:  
\begin{small}
\begin{align}
\frac{m_2}{m_1} e^{2 i \alpha }&= -\frac{\tan \theta _{12} \left(e^{i \delta } \sin \theta _{13}
   \sin 2 \theta _{23} \cos\theta _{12}-\sin \theta _{12} \sin
   ^2\theta _{23}+\sin \theta _{12} \cos ^2\theta _{23}\right)}{\cos
   \theta _{12} \cos 2 \theta _{23}-2 e^{i \delta } \sin \theta
   _{12} \sin \theta _{13} \sin\theta _{23} \cos\theta _{23}} ,\\
\frac{m_3}{m_1} e^{2 i \beta }&= \frac{\tan \theta _{13} \left(e^{i \delta } \cos 2 \theta
   _{12} \cos 2 \theta _{23} \tan \theta _{13} \sec\theta
   _{12}-\sin \theta _{12} \sin 2 \theta _{23} \left(\sec\theta
   _{13}+e^{2 i \delta } \sin \theta _{13} \tan \theta
   _{13}\right)\right)}{\sin\theta _{12} \sin \theta _{13} \sin 2
   \theta _{23}-e^{-i \delta } \cos\theta _{12} \cos 2 \theta _{23}}.
\end{align}
\end{small}
Only two textures belonging to class C viz. $C^{II}$ and $C^{VI}$ are compatible with the neutrino data and have been described below.\\
\textbf{Texture $C^{II}$}: The neutrino mass matrix corresponding to texture $C^{II}$ is given by
\begin{eqnarray}
C^{II}= \left(\begin{array}{ccc}
 \Delta & \times & \Delta \\
 \times & 0 & \times \\
 \Delta & \times & 0 \\
\end{array}
\right).
\end{eqnarray}
Eq. (22) can be used to obtain the following equations for the mixing angle $\theta_{12}$ and unphysical phases ($\phi_e,\phi_{\tau}$) for texture $C^{II}$:
\begin{equation}
\sin 2\theta_{12}=\frac{P}{\sqrt{P^2+Q^2}}~~\textrm{and}~~\sin(\phi_e-\phi_{\tau})=\frac{U^2 W \pm V \sqrt{U^2 \left(16 U^2+16 V^2-W^2\right)}}{U^3+U V^2}
\end{equation}
where
\begin{align}
P&=\frac{1}{2} \sec\theta _{13} (\left(\cos\theta _{23}+\cos 3 \theta
   _{23}\right) \tan\theta _{13} \sin \left(2 \delta +\phi _{\tau }\right)\nonumber \\
   &+2 \cos 2 \theta _{13} \cos 2 \theta _{23} \sec ^2\theta _{13} \sin
   \left(\delta +\phi _e\right)),\nonumber \\
Q&=\sin \theta _{23} \left(4 \cos ^2\theta _{23} \tan ^2\theta _{13}
   \sin \left(3 \delta +\phi _{\tau }\right)+2 \sec ^2\theta _{13} \left(\sin ^2\theta
   _{13}+\cos 2 \theta _{23}\right) \sin \left(\delta +\phi _{\tau }\right)\right)\nonumber \\
   & +\sin 2 \theta _{23} \tan \theta _{13} \left(2 \cos 2 \theta _{13} \sec
   ^2\theta _{13} \sin \left(2 \delta +\phi _e\right)-2 \tan ^2\theta _{13} \sin
   \phi _e\right),\nonumber\\
U&=\cos 2 \delta \sin ^2\theta _{13} \left(\sin 3 \theta _{23}-\sin \theta
   _{23}\right) \cos ^2\theta _{23}+\sin \theta _{23} \cos 2 \theta
   _{13} \cos ^22 \theta _{23}, \\
V&=\sin 2 \delta \sin ^2\theta _{13} \left(\sin 3 \theta _{23}-\sin \theta
   _{23}\right) \cos ^2\theta _{23},\nonumber\\
W&=\sin\delta \left(2 \sin 4 \theta _{23} \cos 2 \theta _{13} \tan\theta
   _{13}-\sin 2 \theta _{13} \sin 2 \theta _{23} \cos ^2 2 \theta
   _{23}\right).\nonumber
\end{align}

\textbf{Texture $C^{VI}$}: The $C^{VI}$ texture neutrino mass matrix is given by
\begin{eqnarray}
C^{VI}= \left(\begin{array}{ccc}
 \times & \times & \Delta \\
 \times & 0 & \Delta \\
 \Delta & \Delta & 0 \\
\end{array}
\right).
\end{eqnarray}
The corresponding mixing angle $\theta_{12}$ and the Dirac phase $\delta$ can be calculated from Eq. (22) to obtain
\begin{equation}
\sin 2\theta_{12}=-\frac{P}{\sqrt{P^2+Q^2}}~~\textrm{and}~~\sin\delta=\tan \theta _{13} \csc \theta _{23} \sec 2 \theta _{23} \sin \left(\phi _e-\phi _{\mu }\right)
\end{equation}
where
\begin{align}
P&= \sin \theta _{13} \left(\cos\theta _{23}+\cos 3 \theta _{23}\right)
   \cos \left(\delta +\phi _e\right),\\
Q&= \frac{1}{2} (\sin 2 \theta _{13} \cos \left(\delta +\phi _{\mu }\right)+\sin \theta
   _{23} (\left(\cos 2 \theta _{13}-2 \cos 2 \theta _{23}-1\right)
   \cos \phi _e\nonumber \\
   &-4 \sin ^2 \theta _{13} \cos ^2 \theta _{23} \cos
   \left(2 \delta +\phi _e\right))).\nonumber
\end{align}

\section{Numerical Results and Discussion}
We have performed a numerical analysis for all the forty-two textures shown in Table I. A set of around $\!\sim\! 10^{7}$-$10^{8}$ random numbers are generated for each unknown parameter. Two mixing angles $(\theta_{13},\theta_{23})$ and the neutrino mass squared differences $(\Delta m_{21}^2,\Delta m_{31}^2$ $(\Delta m_{23}^2)$ for NO(IO)) are generated randomly within their $3\sigma$ experimental ranges (Table II). The parameter $m_1$ is generated freely and the phase angles $(\phi_e,\phi_\mu,\phi_\tau)$ are generated in the range ($0,2\pi$) randomly except for the texture $C^{II}$ where $\delta$ is generated randomly. The parameter $R_\nu$ is calculated by using Eq.(20) and only those values of the required input parameters ($\theta_{13},\theta_{23},m_1,\phi_e,\phi_\mu,\phi_\tau$) are acceptable for which  $R_{\nu}$ falls within the $3\sigma$ experimental range. The mass eigenvalues $m_2, m_3$ can be calculated by using Eq.(19) and the two Majorana-type CP-violating phases $\alpha,\beta$ can be extracted by using Eq.(18). The relations given in Eqs. (26, 29, 32, 37, 40, 43, 48, 51) are used to calculate the mixing angle $\theta_{12}$ and the Dirac-type CP-violating phase $\delta$ for all viable textures except texture $C^{II}$ (where ($\phi_e-\phi_\tau$) is calculated instead of $\delta$). The allowed points of mixing angle $\theta_{12}$ is required to lie within the $3\sigma$ experimental range given in Table II. We present the allowed ranges of various oscillation parameters for each viable pattern in Table III and Figs. 1-6. Textures $A_{1}^{IV}$-$A_{2}^{VI}$, $A_{1}^{V}$-$A_{2}^{V}$ and $A_{1}^{VI}$-$A_{2}^{IV}$ are related by $\mu$-$\tau$ symmetry as shown in Figs. 1, 2 and 3, respectively. Some interesting observations are summarized below:
\begin{itemize}
\item[i).] Among the forty-two possible (twenty-two independent) textures, only eight textures viz. $A_{1}^{IV}, A_{1}^{V},A_{1}^{VI}$, $A_{2}^{IV}, A_{2}^{V}, A_{2}^{VI},C^{II}$ and $C^{VI}$ are phenomenologically viable in the light of current neutrino oscillation data at $3\sigma$ C.L.
\item[ii).] All viable textures follow a specific mass ordering. The textures $A_{1}^{IV}, A_{1}^{V},A_{1}^{VI}$, $A_{2}^{IV}, A_{2}^{V}, A_{2}^{VI}$ are allowed only for NO spectrum whereas textures $C^{II}, C^{VI}$ are allowed only for an inverted mass spectrum.
\item[iii).]Texture $A_{1}^{IV}$ and $A_{1}^{V}$ predicts $\theta_{23}<45^{\circ}$ while for textures $A_{2}^{V}, A_{2}^{VI}, C^{II}$ and $C^{VI}$ we get $\theta_{23}>45^{\circ}$.
\item[iv).] All viable textures with NO spectrum predict vanishing $|M_{ee}|$. On the other hand, textures with IO spectrum $C^{II}$ and $C^{VI}$ predict $|M_{ee}|$ in the range ($0.037$-$0.039$) eV  and ($0.036$-$0.038$) eV, respectively, which is within the reach of the forthcoming experiments.
\item[v).] All the viable textures admit a large Dirac CP-violation leading to a large value of the Jarlskog rephasing invariant $J_{CP}>0.02$. For textures $C^{II}$ and $C^{IV}$, the Dirac-type CP-violating phase $\delta\neq \pi/2$ or $3\pi/2$, whereas for other viable textures $\delta=\pi/2$ or $3\pi/2$ is acceptable.
\end{itemize} 

\begin{table}[h]
\begin{scriptsize}
\begin{center}
\caption{Numerical predictions for all viable textures having two texture zeros and one equality in $M_{\nu}$ at $3\sigma$ C.L.}
\label{table:tm2}
\begin{tabular}{lllllllll}
 \hline\hline
Texture & Spectrum & $m_{lowest}$ (meV)& $\Sigma$ (meV)&$J_{CP}$&$\delta^\circ$&$\alpha^\circ$&$\beta^\circ$&$\theta_{23}^\circ$\\
 \hline
$A_{1}^{IV} $ &  NO &  4.3-5.6 & 63.4-67.5 & $\pm$(0.029-0.036) & 72-110 $\oplus$ & 79-82 $\oplus$  & 31-48 $\oplus$ & 40.3-44.3 \\
&            &     &                 &                 &   (251-287)  &  (278-281) &  (312-329)          & \\
$A_{1}^{V} $ &   NO &  4.8-6.5 & 64.4-68.7 & $-$0.037-0.037 & 69-289  & 80-90 $\oplus$ & 28-90 $\oplus$ & 43.4-43.9 \\
&            &     &                 &                 &  &  (270-280)  &   (270-331)          & \\
$A_{1}^{VI} $ &  NO &  3.7-8.7  & 62.6-73 & $-$0.036-0.036 & 35-325 & 80-90 $\oplus$& 13-90 $\oplus$  & 40.3-51.3\\
&            &     &                 &                 &        &   (270-280) &   (270-347)        & \\
$A_{2}^{IV} $ &  NO &  3.0-8.1  & 61.8-71.7  & $-$0.036-0.036 & 24-338 & 80-90 $\oplus$ & 0-78 $\oplus$ & 40.3-51.2\\
&            &     &                 &                 &     &      (270-280)  &   (282-360)         & \\
$A_{2}^{V} $ &   NO &  4.8-6.5 & 64.2-68.7 & $-$0.036-0.036 & 0-108$\oplus$  & 80-90 $\oplus$ & 29-90 $\oplus$ & 46.1-46.6\\
&            &     &                 &                 &    (250-360) &   (270-280)  &   (270-330)         & \\
$A_{2}^{VI} $ &  NO &  4.2-5.7 & 63.4-67.4 & $\pm$(0.026-0.037) & 56-108 $\oplus$ & 79-82 $\oplus$ & 31-56 $\oplus$ & 45.6-51.8\\
&            &     &                 &                 &     (254-304) &  (278-281) &  (304-330) & \\
$C^{II} $ & IO & 37.5-39.3 & 160.5-167.7 & $\pm$(0.023-0.033) & 114-132 $\oplus$& 56-63 $\oplus$  & 37-50 $\oplus$ & 50.3-50.8\\
&            &     &                 &                 &   (228-246) &  (297-304) &  (310-323) & \\
$C^{VI} $ & IO & 36.6-38.2  &158.5-165.2   & $\pm$(0.022-0.033) & 115-134 $\oplus$ & 57-64 $\oplus$ & 36-49 $\oplus$ & 50.4-51\\
&            &     &                 &                 &   (226-245) &   (296-303) &  (311-324) & \\
 \hline
 \end{tabular}
\end{center}
\end{scriptsize}
\end{table}

\begin{figure}[h]
\begin{center}
\epsfig{file=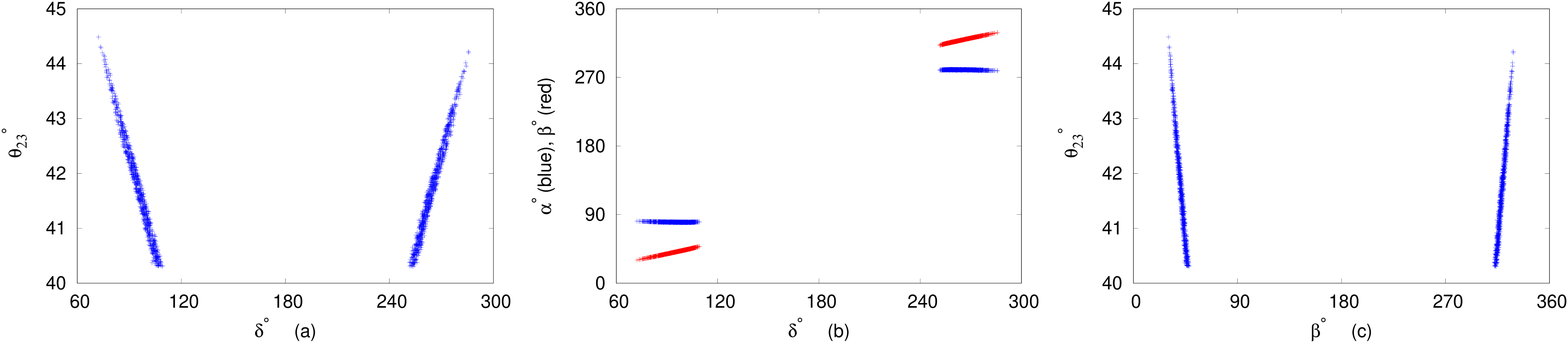, width=15cm, height=5cm}\\
\epsfig{file=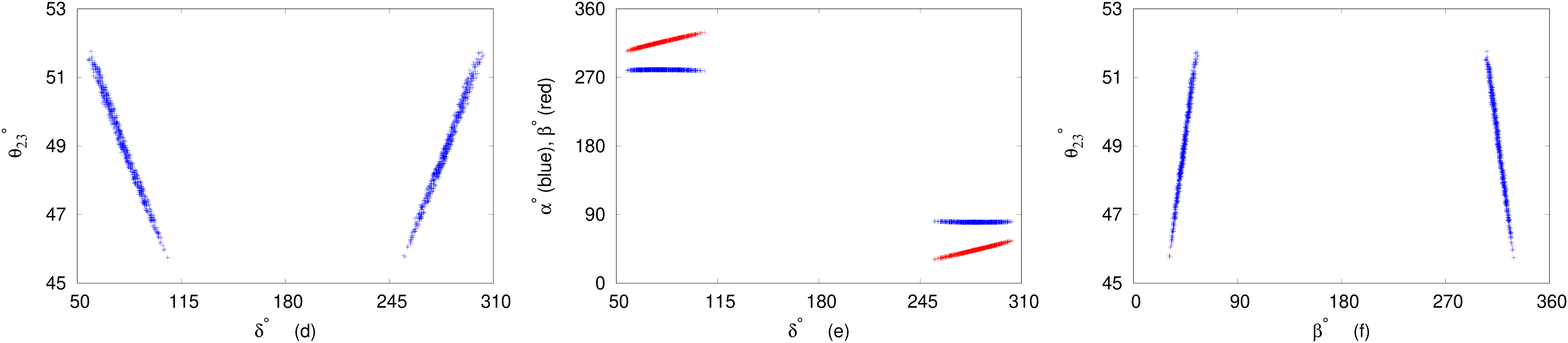, width=15cm, height=5cm}
\end{center}
\caption{Correlation plots among various parameters for $\mu$-$\tau$ symmetric textures $A_{1}^{IV}$(first row) and $A_{2}^{VI}$(second row) with NO.}
\end{figure}

\begin{figure}[h]
\begin{center}
\epsfig{file=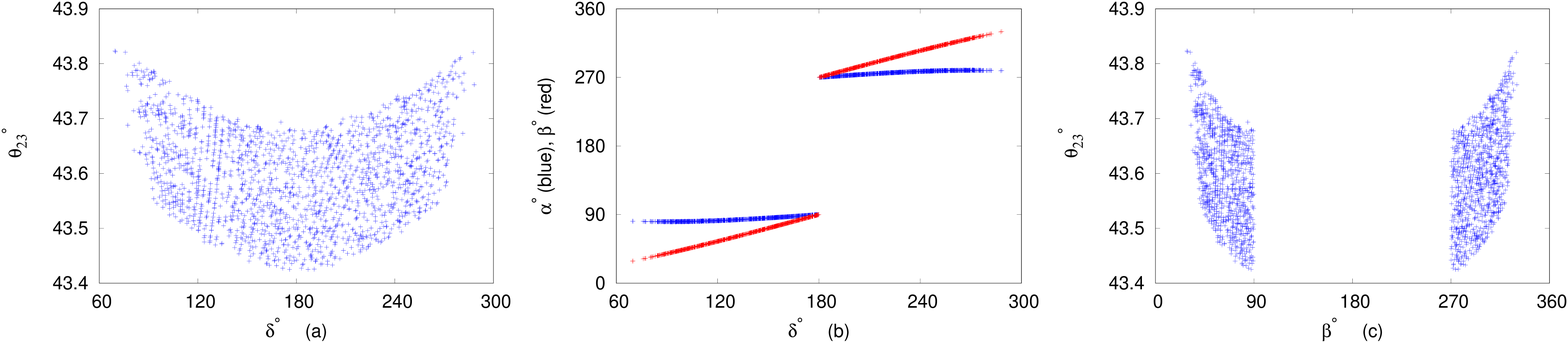, width=15cm, height=5cm}\\
\epsfig{file=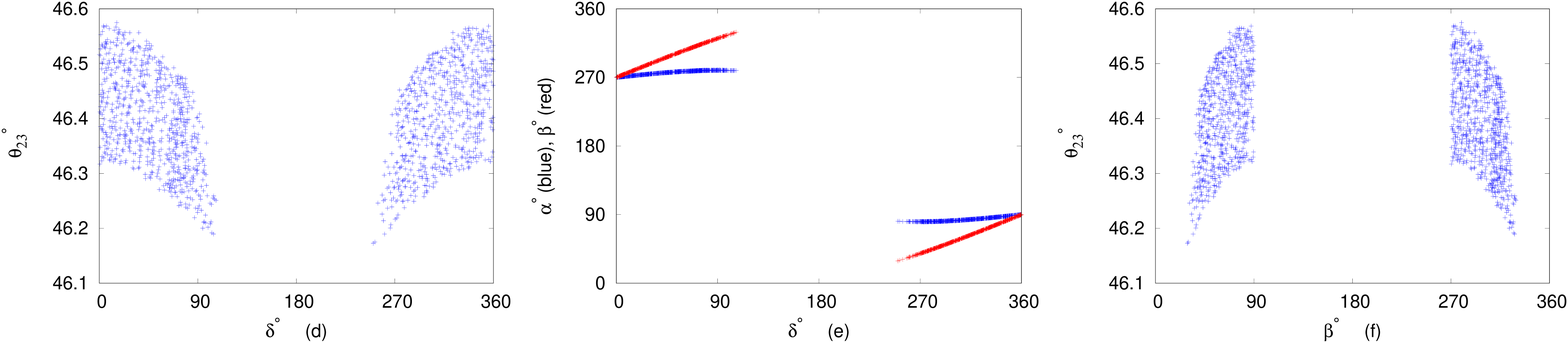, width=15cm, height=5cm}
\end{center}
\caption{Correlation plots among various parameters for $\mu$-$\tau$ symmetric textures $A_{1}^{V}$(first row) and $A_{2}^{V}$(second row) with NO.}
\end{figure}

\begin{figure}[h]
\begin{center}
\epsfig{file=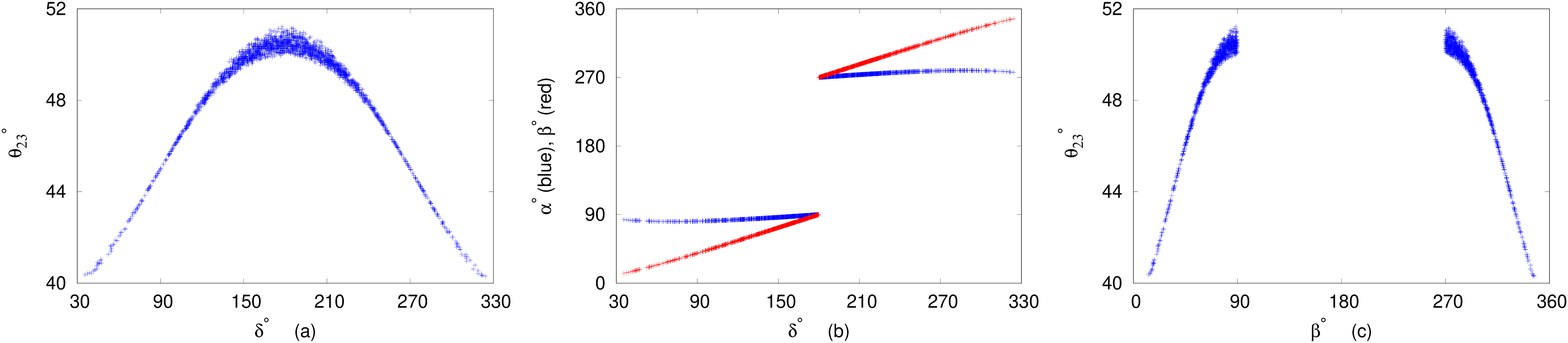, width=15cm, height=5cm}\\
\epsfig{file=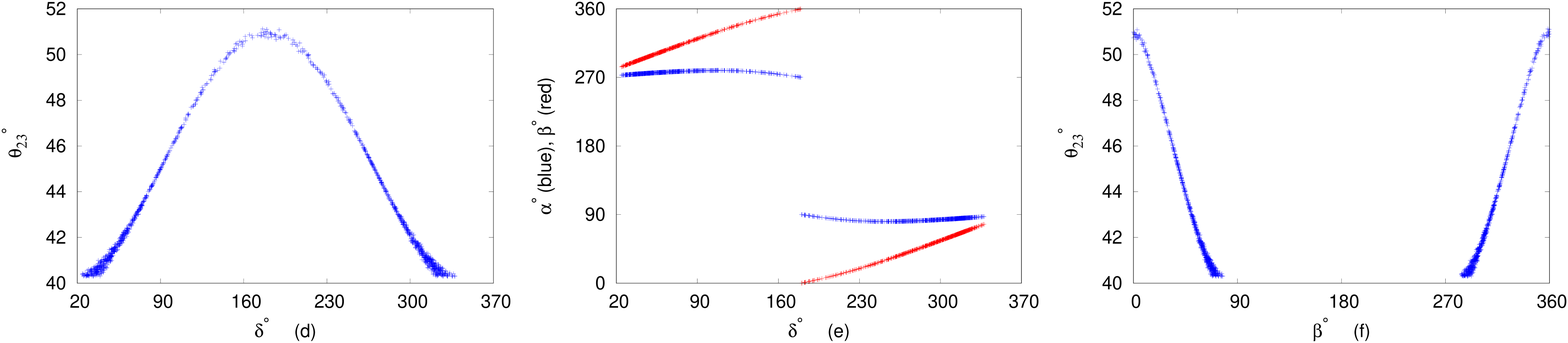, width=15cm, height=5cm}
\end{center}
\caption{Correlation plots among various parameters for $\mu$-$\tau$ symmetric textures $A_{1}^{VI}$(first row) and $A_{2}^{IV}$(second row) with NO.}
\end{figure}

\begin{figure}[h]
\begin{center}
\epsfig{file=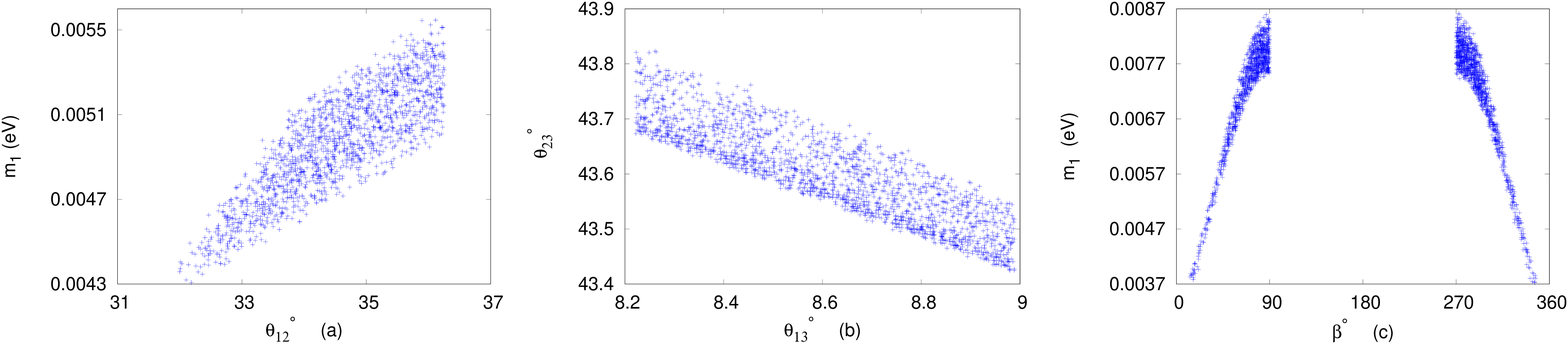, width=15cm, height=5cm}\\
\epsfig{file=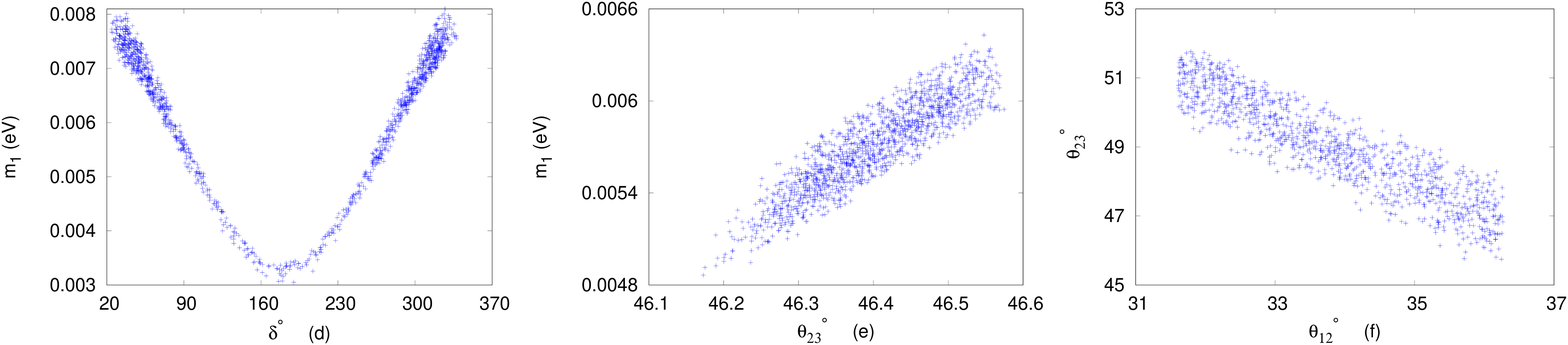, width=15cm, height=5cm}
\end{center}
\caption{Correlation plots for textures $A_{1}^{IV}$(a), $A_{1}^{V}$(b), $A_{1}^{VI}$(c), $A_{2}^{IV}$(d), $A_{2}^{V}$(e) and $A_{2}^{VI}$(f) with NO.}
\end{figure}

\begin{figure}[h]
\begin{center}
\epsfig{file=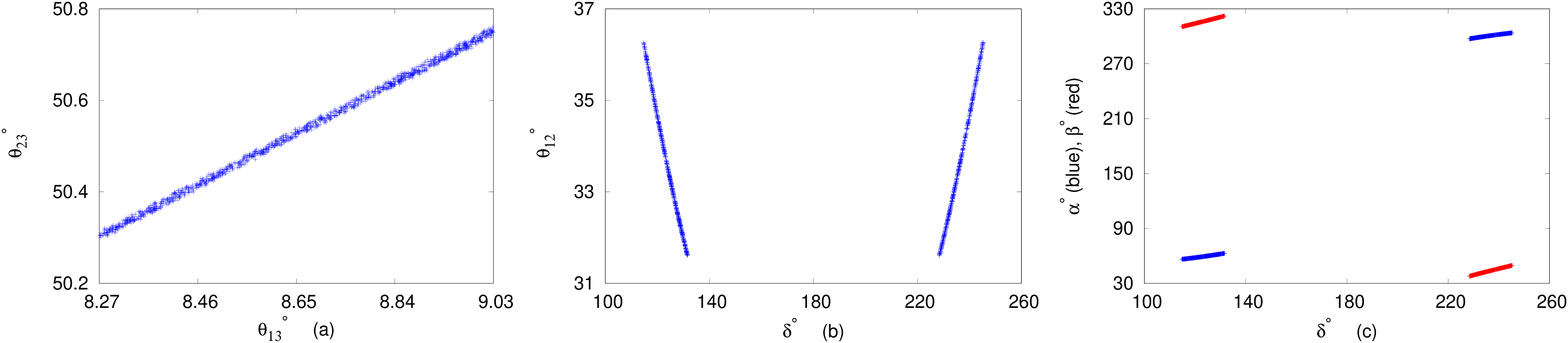, width=15cm, height=5cm}\\
\epsfig{file=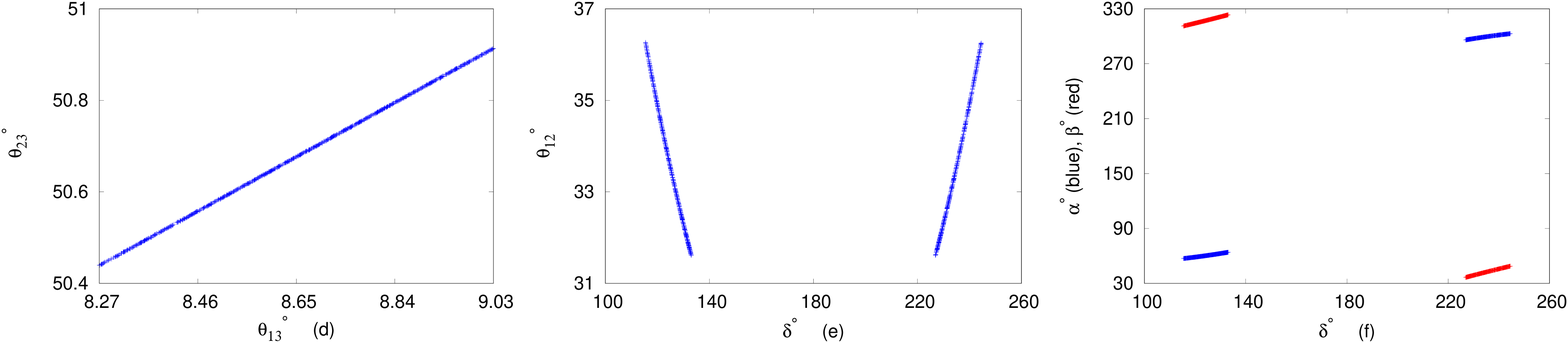, width=15cm, height=5cm}
\end{center}
\caption{Correlation plots among various parameters for textures $C^{II}$(first row) and $C^{VI}$(second row) with IO.}
\end{figure}
\begin{figure}[h]
\begin{center}
\epsfig{file=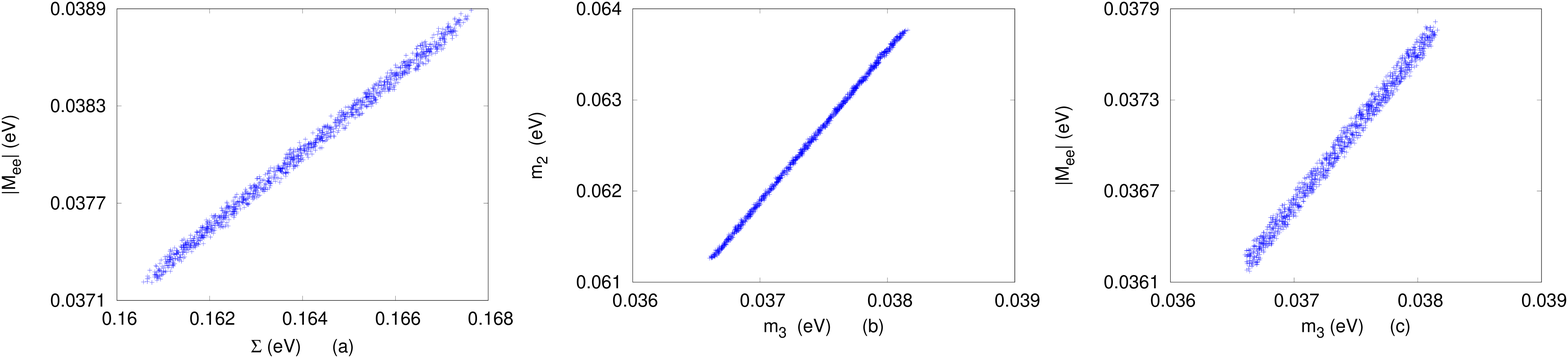, width=15cm, height=5cm}
\end{center}
\caption{Correlation plots for textures $C^{II}$(a) and $C^{VI}$(b,c) with IO.}
\end{figure}

\section{Symmetry Realization}
It has been shown in the literature that the texture zeros or vanishing cofactors in the Majorana neutrino mass matrix can be realized by implementing the $Z_{n}$ flavor symmetries. However, texture equality between two elements cannot be achieved by imposing any Abelian symmetry. Therefore, non-Abelian flavor symmetry is used to realize such textures. In the present work, we realize the hybrid textures having two texture zeros and a texture equality in $M_{\nu}$ within the framework of type-II seesaw mechanism (as disscussed in Ref.\cite{liu}) using $S_3$ symmetry along with $Z_3$ group. The type-II seesaw model extends the scalar sector of the Standard Model (SM) by one or more $SU(2)_{L}$ scalar triplets \cite{seesaw-II}.\\
Here, we present an illustrative simple $S_3\times Z_3$ model to realize one of the viable textures, e.g. texture $A_1^{V}$, studied in this analysis. Within the framework of the type-II seesaw mechanism, atleast five scalar triplets $\Delta_i$ and two scalar doublets $\Phi_i$ are required to realize this texture. The scalar triplets, lepton doublets, right handed charged lepton singlets and scalar doublets under the $Z_3 \times S_3$ symmetry are required to transform as follows:
\begin{align}
& \bar{L}_{eL}\sim (\omega^2,1), ~~\left(\begin{array}{c}  \bar{L}_{\mu L} \\
 \bar{L}_{\tau L} \\ \end{array} \right)\sim (\omega,2),\nonumber \\
 &\left(\begin{array}{c} \Delta_1 \\
 \Delta_2 \\ \end{array} \right)\sim (1,2), ~~~\left(\begin{array}{c} \Delta_3 \\
 \Delta_4 \\ \end{array} \right)\sim (\omega,2), ~~\Delta_5\sim (\omega,1), \nonumber\\
& e_R \sim(\omega,1),~~~ \mu_R \sim (\omega^2,1^\prime), ~~~\tau_R \sim (\omega^2,1),~~~\left(\begin{array}{c} \Phi_1 \\
 \Phi_2 \\ \end{array} \right) \sim (1,2).       
\end{align}
These transformations lead to the following $S_3\times Z_3$ invariant Lagrangian for leptons:
\begin{align}
\mathcal{L}&=Y_1 (\bar{L}_{eL}\Delta_1 L^c_{\mu L}+\bar{L}_{eL}\Delta_2 L^c_{\tau L})+Y_2(\bar{L}_{\mu L}\Delta_5 L^c_{\mu L}+\bar{L}_{\tau L}\Delta_5 L^c_{\tau L})\nonumber \\
&+Y_3[(\bar{L}_{\mu L}\Delta_3 L^c_{\tau L}+\bar{L}_{\tau L}\Delta_3 L^c_{\mu L})+( \bar{L}_{\mu L}\Delta_4 L^c_{\mu L}- \bar{L}_{\tau L}\Delta_4 L^c_{\tau L})]\nonumber\\
&+Y_e \bar{L}_{eL}H e_R+Y_{\mu}(\bar{L}_{\mu L}\Phi_2 \mu_R-\bar{L}_{\tau L}\Phi_1 \mu_R)+Y_{\tau}(\bar{L}_{\mu L}\Phi_1 \tau_R+\bar{L}_{\tau L}\Phi_2 \tau_R)+h.c.
\end{align}
where $H\sim(1,1)$ is the SM Higgs doublet. When three scalar doublets acquire vacuum exception values (VEVs) $\langle H\rangle=v/\sqrt{2},\langle\Phi_2\rangle=u/\sqrt{2}$ and $\langle\Phi_1\rangle=0$, the $S_3\times Z_3$ invariant Yukawa Lagrangian leads to a diagonal charged lepton mass matrix
\begin{equation}
M_l=\left(\begin{array}{ccc} m_e & 0 & 0 \\
 0 & m_\mu & 0 \\
 0 & 0& m_\tau \\ \end{array} \right)
\end{equation}
with $m_e=Y_e v/\sqrt{2}, m_\mu=Y_\mu u/\sqrt{2}$ and $m_\tau=Y_\tau u/\sqrt{2}$. In neutrino sector, by choosing the VEV's of scalar triplets $\langle\Delta_2\rangle=v_2, \langle\Delta_3\rangle=v_3, \langle\Delta_5\rangle=v_5$ and $\langle\Delta_1\rangle=\langle\Delta_4\rangle=0$, we arrive at the following neutrino mass matrix 
\begin{equation}
M_\nu=\left(\begin{array}{ccc} 0 & 0 & \times \\
 0 & \Delta & \times \\
 \times & \times & \Delta\\ \end{array} \right).
\end{equation}
The above mass matrix have two texture zeros at the $(1,1),(1,2)$-positions and a texture equality at the $(2,2),(3,3)$-positions.\\
Although, the $S_3\times Z_3$ model successfully realizes the $A_1^{V}$ texture of $M_\nu$, the main issue in this model is the vacuum alignment problem viz. the scalar fields acquiring specific form of VEVs.\\
Therefore, the symmetry realization for all viable textures for neutrino mass matrices in a systematic and self-consistent way deserves further research

\section{Summary}
In this work, we presented a detailed phenomenological analysis of neutrino mass matrices having two texture zeros and a texture equality between the elements of the neutrino mass matrix. These texture structures can arise through the type-II seesaw mechanism. Out of the twenty-two independent possible textures for $M_{\nu}$, only eight are allowed by present neutrino oscillation data at $3\sigma$ C.L. Textures $A_1^{IV}, A_1^{V}, A_1^{VI}, A_2^{IV}, A_2^{V}$ and $A_2^{VI}$ are allowed only for NO and predict vanishing $|M_{ee}|$ whereas textures $C^{II}$ and $C^{VI}$ are viable for IO only and predict $|M_{ee}|$ of the order of $0.01$ eV which can be probed in the forthcoming neutrinoless double beta decay experiments. All viable textures are fairly predictive and predict interesting correlations among various neutrino oscillation parameters. In addition, there are predictions for the octant of atmospheric mixing angle $\theta_{23}$ and quadrant of the CP-violating phases ($\alpha,\beta$ and $\delta$). To illustrate how such texture structures can be realised, we presented a simple $S_3\times Z_3$ model for the texture structure $A_1^{V}$. As most of the textures studied in this analysis have strong predictions for one or more presently unknown neutrino parameters, experimental results on these neutrino parameters  will help in deciding the form of the neutrino mass matrix.

\section{Acknowledgements}
The research work of S. D. is supported by the Council of Scientific and Industrial Research (CSIR), Government of India, New Delhi vide grant No. 03(1333)/15/EMR-II. S. D. gratefully acknowledges the kind hospitality provided by IUCAA, Pune. Authors thank Radha Raman Gautam and Lal Singh for carefully reading the manuscript.

\end{document}